\documentclass[conference, 10pt]{IEEEtran}

\usepackage{bbm}
\usepackage{bbold}
\usepackage{amsmath}
\usepackage{amssymb}
\usepackage[dvips]{graphicx}
\usepackage[caption = false]{subfig}
\usepackage{graphicx}
\usepackage{cite}
\usepackage{epsfig}
\usepackage{bigints}
\usepackage{stfloats}
\usepackage{amsthm}
\usepackage{float}
\usepackage{bbold}
\usepackage{bbm, dsfont}
\usepackage{xcolor}
\usepackage{multirow}
\usepackage[linesnumbered,ruled,vlined]{algorithm2e}

\theoremstyle{definition}

\let\oldemptyset\emptyset

\IEEEoverridecommandlockouts

\begin{document}

\title{ \vspace{-1cm} Learning-based Caching in Cloud-Aided Wireless Networks \vspace{-1.5ex}}
\author{	
	\IEEEauthorblockN{
	Syed Tamoor-ul-Hassan\IEEEauthorrefmark{1}, Sumudu Samarakoon\IEEEauthorrefmark{1}, Mehdi Bennis\IEEEauthorrefmark{1}, Matti Latva-aho\IEEEauthorrefmark{1}, Choong-Seong Hong\IEEEauthorrefmark{2}}
	\IEEEauthorblockA{\IEEEauthorrefmark{1}Center for Wireless Communications, University of Oulu, Finland, \\
		email: \{tsyed, bennis, sumudu, matti.latva-aho\}@ee.oulu.fi}
	\IEEEauthorblockA{\IEEEauthorrefmark{2}Department of Computer Engineering, Kyung Hee University, South Korea,	email: cshong@khu.ac.kr}
	\\[-7.0ex] 
 }
\vspace{-4ex}
\providecommand{\keywords}[1]{\textbf{\textit{Index terms---}} #1}
\maketitle
\vspace{-5.5cm}
\begin{abstract} 
This paper studies content caching in cloud-aided wireless networks where small cell base stations with limited storage are connected to the cloud via limited capacity fronthaul links. By formulating a utility (inverse of service delay) maximization problem, we propose a cache update algorithm based on spatio-temporal traffic demands. 
To account for the large number of contents, we propose a content clustering algorithm to group similar contents. Subsequently, with the aid of regret learning at small cell base stations and the cloud, each base station caches contents based on the learned content popularity subject to its storage constraints. The performance of the proposed caching algorithm is evaluated for sparse and dense environments while investigating the tradeoff between global and local class popularity. Simulation results show 15\% and 40\% gains in the proposed method compared to various baselines.
\end{abstract}
\vspace{-0.2cm}
\section{Introduction}
\label{sec:Intro}
\vspace{-0.2cm}
Edge caching represents a viable solution to overcome challenges associated with network densification by intelligently caching contents at the network edge \cite{Ref3_Intro}. Besides reducing  latency, edge caching also offloads the backhaul traffic load \cite{Ref5_Intro}. Existing literature investigates the potential benefits of caching in terms of backhaul offloading gains and latency \cite{Ref6_Intro, Ref7_Intro, Ref9_Intro}. While these works show improved network performance through caching, they neglect the intrinsic user behavior by considering a fixed caching policy. Due to the spatio-temporal requests, small cell base stations (SBSs) often need to update their cache following their local request distribution to minimize latency \cite{Ref10_Intro}. In such scenarios, optimal content placement becomes a challenging and non-trivial problem. To serve user requests, the works in \cite{Ref12_Intro, Ref13_Intro, Ref14_Intro} proposed dynamic caching algorithms based on fixed content popularity. However, these works assume a small content library with fixed content popularities. With the growing library size, determining popularity and content caching becomes computationally expensive. Recently, grouping contents based on their popularity was proposed in \cite{Ref18_Intro}. However, how to group contents and cache accordingly based on time-varying popularity was not studied. \\
\indent The main contribution of this paper is to revisit the fundamental problem of content caching under spatio-temporal traffic demands in cloud-aided wireless networks and explore the tradeoffs between global and local content popularity. By considering a random deployment of SBSs and users, the objective is to determine what contents need to be cached locally by every SBS so as to maximize the cache hit rate. Based on the instantaneous content requests, each SBS locally learns the time-varying content popularity with the aid of regret learning \cite{Ref1:Reinforcement_Learning}. Simultaneously, the cloud learns the global content popularity. By randomizing its caching strategy, each SBS optimizes the caching policy in a decentralized manner and updates its cache.
\vspace{-0.3cm}
\section{System Model and Problem Formulation}

\begin{figure}[t]%
\centering
\includegraphics[width = 5.75cm, height = 5.00cm]{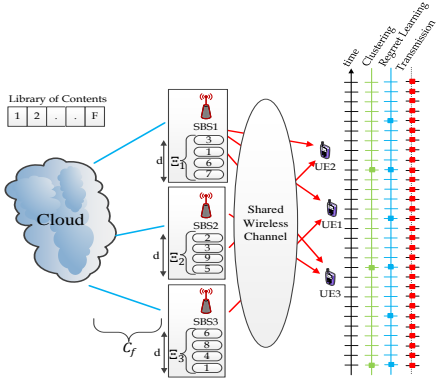}
\vspace{-0.3cm}
\caption{System Model}
\vspace{-0.2cm}
\label{fig:Sys_Model}
\end{figure}

\vspace{-0.1cm}
Consider the downlink transmission of a small cell network comprised of randomly deployed SBSs, $\mathcal{S} = \{1, ..., S\}$ with intensity $\lambda_{\mathrm{SBS}}$. Let $Y_s$ represent the location of the $s$-th SBS. Each SBS serves a set of user equipment(s) (UEs), $\mathcal{U} = \{1, ..., U\}$, deployed randomly with intensity $\lambda_{\mathrm{UE}}$. The location of the $u$-th UE is denoted by $Z_u$. Each SBS serves UEs' requests over a common pool of spectrum with bandwidth $\omega$. Accordingly, the instantaneous data rate of UE $u$ served by SBS $s$ is:

\small
\vspace{-0.5cm}
\begin{equation}
R_{su}(t) = \omega \mathrm{log}_2 \left(1 + \frac {p_{s} \lVert \Psi_{su}(t) \rVert^2} { \sigma^2 + \sum_{s' \in \mathcal{S} \setminus s} p_{s'} \lVert \Psi_{s'u}(t) \rVert^2 }  \right),
\vspace{-0.3cm}
\end{equation}
\normalsize

\noindent where $\sigma^2$ represents the variance of noise, $p_{s}$ denotes the transmit power of SBS $s$ and $\Psi_{su}(t)$ denotes the channel gain between UE $u$ and SBS $s$. \\
\indent Each SBS is equipped with a cache of size $d$ where it stores contents from a content library $\mathcal{F} = \{1, ..., F\}$ as shown in Fig. \ref{fig:Sys_Model}. Let $1/\mu$ be the size of all contents. In addition, let $\boldsymbol{\Xi}(t) = [\boldsymbol{\Xi}_s(t)]_{s \in \mathcal{S}}$ represent the vector of SBSs cache at time $t$ where $\boldsymbol{\Xi}_{s}(t) \subseteq \mathcal{F}$ represents the contents cached by SBS $s \in \mathcal{S}$ at time $t$ such that $|\boldsymbol{\Xi}_{s}(t)| \leq d$. We assume that SBSs partition the content library into popularity classes such that each content in a class is equally popular i.e., \textit{multi-class model} \cite{Ref18_Intro}. Let the set of contents be partitioned into popularity classes $\mathcal{K} = \{1, 2, ..., K\}$, where $\mathcal{F}_k = \{1, ..., F_k \}$ such that $\mathcal{F}_k \subset \mathcal{F}$, $\mathcal{F}_k \cap \mathcal{F}_{k'} = \oldemptyset$, $k \neq k'$. Due to the constrained cache size and lack of coordination among SBSs, each SBS is connected to the cloud via a fixed capacity fronthaul link $C_f$ to obtain the global content popularity and update its cache accordingly. \\
\indent Each UE requests contents from the library following the dynamic popularity model i.e., spatio-temporal model\cite{Ref1_UEReq}. Let the content demanded by the $u$-th UE at time $t$ is denoted by $q_u(t)$ such that $q_u(t) \in \{0, 1, 2, ..., F\}$ where $q_u(t) = 0$ denotes no request by user $u$ at time $t$. For simplicity, we assume that each UE requests one content at a time. Let the content demand vector at SBS $s$ be $\boldsymbol{D}_s(t) = [D_{sf}(t)]_{f \in \mathcal{F}}$ such that $D_{sf}(t) = \sum_{u \in \mathcal{N}_s} \mathbb{1}_{q_u(t) = f}$ where $\mathbb{1}_{x}$ is the indicator function and $\mathcal{N}_s$ denotes the users in the coverage of SBS $s$. \\
\indent The instantaneous reward of a SBS depends on the instantaneous cache hits and service rate. Absence of a requested content from a SBS incurs a cache miss. If the content is cached by multiple SBSs in UE's coverage, the user associates to the nearest SBS caching the requested content. In this regard, the reward of SBS $s$ for serving UE $u$ is given by:
\vspace{-0.35cm}
\begin{equation}
\label{eq:reward}
g_{sq_u}(t, \boldsymbol{\Xi}_s(t)) = \mathbb{1}_{\{q_u(t) \in \boldsymbol{\Xi}_s(t)\}} R_{su}(t).
\end{equation}
\vspace{-0.75cm}
\subsection{Utility Maximization Problem}
\label{ssec: Prob_Form}
\vspace{-0.1cm}
The objective of SBSs is to determine a caching policy that maximizes their reward while ensuring UEs QoS. From \eqref{eq:reward}, it can be observed that the reward of a SBS depends on the achievable rate and caching policy, i.e., SBS is rewarded if and only if it caches the requested content. For simplicity, fronthaul links are assumed to be used only for cache update and service rate is considered to be zero if the SBS has not cached the content. One of the challenges associated with cache update is when the number of most popular contents is larger than the cache size. In this case, SBSs must update their caching decisions carefully as caching less popular contents may decrease the SBS's reward. \\
\indent For a few UE requests, content popularity at SBSs may not be determined accurately, resulting in poor caching policy yielding lower reward. Hence, it is important that enough statistics are available to better learn the content popularity. To overcome this issue, the cloud estimates the global content popularity gathered from all SBSs. However, acquiring global demand and cache update incurs additional cost given by:
\vspace{-0.25cm} 
\begin{equation}
\varepsilon_s = 1 - \frac {\tau_s} {T_2}
\vspace{-0.25cm}
\end{equation}
where $\tau_s < T_2$ is the time required for cache update and $T_2$ represents the time during which the users' requests are observed. Assume $C_s$ is the fronthaul capacity for SBS $s$, the time required to update the cache of SBS $s$ is:
\vspace{-0.2cm}
\begin{equation}
\tau_{s} = l_p \frac {N} {\mu C_s}, 
\vspace{-0.2cm}
\end{equation} 
where $l_p > 0$ is a constant and $N$ represents the number of new contents. Let $\Xi_s$ be the vector of caching policies at SBS s over time $t = \{0,1,2,...\}$, i.e. $\Xi_s = [\Xi(0), \Xi(1), \Xi(2), ...]$ and $\bar{g}_{su}(\boldsymbol{\Xi}_s) =  \mathrm{lim}_{t \to \infty} \frac {1} {t} \sum_{\tau = 0}^{t-1} g_{sq_u}(\tau, \boldsymbol{\Xi}_s(\tau))$ be the limiting time-average reward at SBS $s$.Then, for each SBS $s \in \mathcal{S}$, the average per-SBS utility is the aggregate utility of the associated UEs i.e., 
\vspace{-0.3cm}
\begin{equation}
\label{eq:UtilityFunction} 
\Upsilon_s(\boldsymbol{\Xi}_s) = \varepsilon_s \sum_{\forall u \in \mathcal{N}_s} \Upsilon_{su}(\bar{g}_{su}(\boldsymbol{\Xi}_s)),
\vspace{-0.3cm} 
\end{equation}
where $\Upsilon_{su}(\bar{g}_{su}) = (1/\mu)^{-1} \bar{g}_{su}$. The network utility maximization problem is:

\vspace{-0.7cm}

\begin{subequations}
\label{eq: Obj_Function}
\centering
\begin{align}
    \underset{\boldsymbol{\Xi}_s}{\text{maximize}} \ \ \ & \textstyle \sum \limits_{\forall s \in \mathcal{S}} \Upsilon_s(\boldsymbol{\Xi}_s) \label{eq:const} \\
    \text{subject to} \ \ & |\boldsymbol{\Xi}_s| \leq d, \ \forall s \in \mathcal{S} \label{eq:const3} \\		
		\ \ & g_{sq_u}(t) > g_{\mathrm{min}}, \ \forall u \in \mathcal{U}, \forall t \label{eq:const4} \\
		\ \ & \textstyle \sum_{\forall s \in \mathcal{S}} C_s \leq C_f, \ \label{eq:const6} \\
		\ \ |Y_{s(u)} - Z_u| < & |Y_{s'} - Z_u|, \forall u \in \mathcal{U}, s' \in \mathcal{S} \setminus s(u) \label{eq:const5},		
\end{align}
\end{subequations}
\vspace{-0.8cm}
\normalsize

where \eqref{eq:const4} is the minimum QoS threshold, \eqref{eq:const5} represents a nearest UE-SBS association, \eqref{eq:const6} is the fronthaul capacity constraint and $s(u)$ represents the serving SBS of user $u$.
\vspace{-0.3cm}
\section{Demand-Based Content Clustering}
\vspace{-0.1cm}
The problem in \eqref{eq: Obj_Function} is trivial when $F \approx d$. With the increasing library size, the problem becomes non-trivial. Moreover, due to time-varying popularity of contents, the complexity of \eqref{eq: Obj_Function} increases manifolds, making the problem extremely challenging to solve. It has been observed that in a real system, there exists a correlation among contents requests i.e., request of a content is nearly similar to one or more contents \cite{Ref7_Intro}.  
This suggests grouping contents based on their demands as a solution to improve caching decisions. By observing the content demand over a finite time period, contents are clustered into different classes where contents in the same class have similar popularity. Thus, \eqref{eq: Obj_Function} is solved over classes rather than contents. In this work, a similarity measure between demand vectors is used to cluster contents into classes. Since, content similarity varies slowly over time, content clustering is a slower process than cache update. In other words, the content clustering remains fixed for a period $T_1 > T_2$ where $T_2$ is the cache update time. To define the similarity measure, let $\boldsymbol{\mathrm{M}}(t) = [\boldsymbol{M}_{ff'}(t)]_{f,f' \in \mathcal{F}}$ be the similarity matrix at time $t$ with:
\vspace{-0.5cm}

\footnotesize
\begin{equation}
\label{eq:Similarity}
\boldsymbol{M}_{ff'}(t) = \exp \left(- \frac {|D_{f}(t) \pi_f(t) - D_{f'}(t) \pi_{f'}(t)|^2} {2 \sigma_l^2} \right) \ \ \ \forall f' \in \mathcal{F},
\end{equation}
\vspace{-0.4cm}
\normalsize

\noindent where $\pi_{f}(t)$ is the popularity of file $f$ at time $t$, $D_f(t)$ is the instantaneous request of file $f$ and $\sigma_l^2$ controls the impact of popularity on similarity. In order to find the content demand vector over the network, all the SBSs upload 
their demand vectors $\boldsymbol{D}_{s}(t)$ to the cloud which computes the network wide demand vector $\boldsymbol{\mathrm{D}}'(t) = \sum_{s \in \mathcal{S}} \boldsymbol{D}_s(t)$ and broadcasts it to all SBSs. Thereafter, SBSs perform content clustering based on the following demand vector:
\vspace{-0.4cm}

\begin{equation}
\label{eq:Global_demand}
\mathrm{\boldsymbol{D}}_s^{'}(t) = \alpha \boldsymbol{\mathrm{D}}'(t) + (1 - \alpha) \boldsymbol{D}_s(t) \ \ 0 \leq \alpha \leq 1,
\end{equation}
\vspace{-0.6cm}

\noindent where $\alpha$ is a tunable parameter that captures local vs. global demand. In this work, spectral clustering technique is used to perform content clustering \cite{Ref1:Clustering1} that exploits the frequency of content requests from the users in the coverage of SBSs and the variance of the similarity matrix to form content classes. The content clustering algorithm at each SBS is explained in Algorithm \ref{algo:algo1}.    
\LinesNumberedHidden{
\begin{algorithm}[t]
\footnotesize
\caption{Content Clustering and Cache Update}
\label{algo:algo1}
\DontPrintSemicolon 
\textbf{Input}: Observed local content demand vector $\boldsymbol{D}_s(t)$ and Global/local tradeoff parameter $\beta$. \\
\KwResult{Content cluster at SBSs $\mathcal{K}_s = \{1, ... K_s\}$, $\forall s \in \mathcal{S}$.}

\textbf{Algorithm:}

\textbf{Phase I - Similarity Matrix Computation;}
\begin{itemize}
     \item Transmit the local demand vector $\boldsymbol{D}_s(t)$ to the cloud.
     \item Compute the similarity matrix $\boldsymbol{\mathrm{M}}(t)$ based on \eqref{eq:Similarity}.
\end{itemize}		
		
\textbf{Phase II - Spectral Clustering Algorithm;}
\begin{itemize}
		\item Compute the diagonal degree matrix $\boldsymbol{X}$ where $X_i = \sum_{\forall f \in \mathcal{F}} m_{ij}$.
    \item Compute the graph laplacian matrix $\boldsymbol{L} = \boldsymbol{X} - \boldsymbol{\mathrm{M}}(t)$.		
    \item Normalize the graph laplacian matrix $\boldsymbol{L}_{\mathrm{norm}} = \boldsymbol{X}^{- \frac {1} {2}}\boldsymbol{L}	\boldsymbol{X}^{\frac {1} {2}}$.
		\item Select a number of $k_{\mathrm{max}}$ eigenvalues of $\boldsymbol{L}_{\mathrm{norm}}$ such that $\lambda_1 \leq ... \leq \lambda_{i_{\mathrm{max}}}$ where $k_{\mathrm{max}}$ is the maximum number of clusters and $\lambda_i$ is the $i-th$ smallest value of $\boldsymbol{L}$.
		\item Choose $k = \mathrm{max}_{i = k_{\mathrm{min}}, ..., k_{\mathrm{max}}} \Delta_i$ where $\Delta_i = \lambda_{i+1} - \lambda_i$.
		\item Calculate $k$ smallest eigenvectors and apply $k$-means clustering to cluster rows of eigenvectors.
\end{itemize}

\textbf{Phase III - Regret Learning and Cache Update;}
\begin{itemize}
		\item Each SBS learns the probability distribution vector $\boldsymbol{\pi}_{s}$ based on \eqref{eq:Regret}.
    \item The cloud learns the probability distribution vector $\boldsymbol{\pi}_c$ based on \eqref{eq:Regret}.
		\item Each SBS updates its cache based on the mixed distribution $\boldsymbol{\pi}' = (1-\beta)\boldsymbol{\pi}_{s} + \beta \boldsymbol{\pi}_c$.
\end{itemize}
\end{algorithm}
} 
\vspace{-0.3cm} 

\section{Caching via Reinforcement Learning}
\vspace{-0.1cm}
The main objective of an efficient caching strategy is to maximize the cache hits while minimizing the service delay and fronthaul cost. However, designing an efficient caching strategy is extremely challenging without a prior knowledge of user demands. Since the demand vector at each SBS varies from other SBSs due to their spatial location, it is necessary to devise adaptive decentralized algorithms to determine the caching strategy. In this respect, each SBS leverages reinforcement learning (RL) to accurately estimate the caching strategy that maximizes the payoff. \\
\indent To employ RL, each SBS implicitly learns the class popularity based on instantaneous user demands. As per \eqref{eq:const3}, the SBSs cache a subset of library contents. At each time, the SBS determines the set of library content to cache which defines the actions of SBSs. Hence, the action space comprises of caching content/contents of class/classes. Let $\mathcal{A}_s$ denotes the action space of SBS $s$ where $\mathcal{A}_s = [\boldsymbol{\Xi}_s^{k_s}]_{k_s \in \mathcal{K}_s}$ where $\mathcal{K}_s$ represents the set of popularity classes at SBS $s$. Here, the action $\Xi_s^{k_s}=1$ indicates that SBS $s$ caches content(s) of class $k_s$. Thus \eqref{eq:UtilityFunction} can be rewritten as:
\vspace{-0.15cm}
\begin{equation}
\label{eq:Utility_Action}
\vspace{-0.1cm}
\Upsilon_s(\boldsymbol{\Xi}_s^{k_{s}}) = \varepsilon_s \sum_{\forall u \in s(u)} \Upsilon_{su}(\boldsymbol{\Xi}_s^{k_{s}}).
\vspace{-0.2cm}
\end{equation} 
\vspace{-0.05cm}

\indent Since, the requests of users change over time, it is necessary to adapt the caching strategy accordingly. As a result, the caching decision corresponding to content(s) of a class becomes a random variable. Let the probability distribution of the caching strategy at SBS $s$ be $\boldsymbol{\pi}_s(t) = [\pi_{s, \Xi_s^1}(t), ..., \pi_{s, \Xi_s^{k_s}}(t) ]$ where $\pi_{s, \Xi_s^{k_s}}(t) = \mathbb{P}(\Xi_s(t) = \boldsymbol{\Xi}_s^{k_s})$ such that $\sum_{\boldsymbol{\Xi}_s^{k_s} \in \mathcal{A}_s}\pi_{s, \boldsymbol{\Xi}_s^{k_s}}(t) = 1$. \\ 
\indent Let $\tilde{\boldsymbol{\Upsilon}}_{s}(t) = (\tilde{\Upsilon}_{s,\Xi_s^1}(t), ..., \tilde{\Upsilon}_{s,\Xi_s^{|\mathcal{K}_s|}}(t))$ denote the vector of estimated utilities for all actions of SBS $s$ where $\tilde{\Upsilon}_{s,\Xi_s^{k_s}}(t)$ is the estimated utility for action $\Xi_s^{k_s}$ at time $t$. Further, let $\hat{\Upsilon}_{s}(t)$ be the feedback of the utilities from all associated users. Due to the time-varying content demands, each SBS needs to update its cache to maximize the utility. For this, each SBS uses regret learning mechanism to determine the caching strategy. The regret learning mechanism iteratively allows players to explores all possible actions and learn optimal strategies \cite{Ref1:Reinforcement_Learning}. As a result, the main objective of utility maximization recast as a regret minimization problem. Here, the objective is to exploit the actions that yield higher rewards while exploring other actions with lower regrets. This behavior is captured by the Boltzmann-Gibbs (BG) distribution given as \cite{Ref12_Intro}:

\footnotesize
\vspace{-0.4cm}
\begin{equation}
G_{s,\boldsymbol{\Xi}_s^{k_s}}(\tilde{\boldsymbol{r}}_s(t)) = \frac {\exp(\frac {1} {\xi_s} \tilde{r}_{s,\boldsymbol{\Xi}_s^{k_s}}^+(t))} {\sum_{\forall \boldsymbol{\Xi}_s^{'} \in \mathcal{A}_s} \exp(\frac {1} {\xi_s} \tilde{r}_{s,\boldsymbol{\Xi}_s^{'}}^+(t))}, \ \forall \boldsymbol{\Xi}_s^{k_s} \in \mathcal{A}_s, 
\end{equation}
\normalsize
\vspace{-0.3cm}

\noindent where $\xi_s > 0$ is a temperature coefficient, and $\tilde{r}_{s,\boldsymbol{\Xi}_s^{k_s}}^+(t) = \mathrm{max}(0, \tilde{r}_{s,\boldsymbol{\Xi}_s^{k_s}}(t))$. A small value of $\xi_s$ maximizes the sum of regrets which results in a mixed strategy where SBSs expolits the actions with higher regrets at time period $t$. On the contrary, a higher value of $\xi_s$ results in uniform distribution over the action set. At each time instant, the estimation of the utility, regret and probability distribution over the action space $\mathcal{A}_s, \forall s \in \mathcal{S}$ is given as:

\footnotesize
\vspace{-0.7cm}

\begin{align}
\label{eq:Regret}
\tilde{{\Upsilon}}_{s,\Xi_s^{k_s}}(t) &= \tilde{\Upsilon}_{s,\Xi_s^{k_s}}(t-1) + \Gamma_s^1(t) \mathbb{1}_{\{\Xi_s(t) = \Xi_s^{k_s}\}} \bigg[ \hat{\Upsilon}_{s}(t) - \tilde{\Upsilon}_{s,\Xi_s^{k_s}}(t-1) \bigg] \nonumber \\
\vspace{-0.4cm}
\tilde{r}_{s,\Xi_s^{k_s}}(t) &= \tilde{r}_{s,\Xi_s^{k_s}}(t-1) + \Gamma_s^2(t) \bigg( \tilde{\Upsilon}_{s,\Xi_s^{k_s}}(t) - \hat{\Upsilon}_{s}(t) - \nonumber \\
\vspace{-0.4cm}
& \ \ \ \tilde{r}_{s,\Xi_s^{k_s}}(t-1) \bigg) \\
\pi_{s,\Xi_s^{k_s}}(t) &= \pi_{s,\Xi_s^{k_s}}(t-1) + \Gamma_s^3(t) \bigg( G_{s,\Xi_s^{k_s}}(\tilde{\boldsymbol{r}}_s(t)) - \pi_{s,\Xi_s^{k_s}}(t-1) \bigg), \nonumber
\end{align}
\normalsize
\vspace{-0.5cm}

\noindent where the learning rates $\Gamma_s^i(t) \forall i \in \{1, 2, 3\}$ satisfy \cite{Ref1:Reinforcement_Learning}:
\vspace{-0.6cm}
\footnotesize

\begin{align}
\label{eq:Lean_Eq}
(i) \ \mathrm{lim}_{t \to \infty} \sum_{\tau = 1}^{t} \Gamma_s^i(\tau) = +\infty, \ \ \ & \mathrm{lim}_{t \to \infty} \sum_{\tau = 1}^{t} (\Gamma_s^i(\tau))^2 < +\infty \nonumber \\
(ii) \ \mathrm{lim}_{t \to \infty} \sum_{\tau = 1}^{t} \frac {\Gamma_s^2(t)} {\Gamma_s^1(t)} = 0, \ \ \ & \mathrm{lim}_{t \to \infty} \sum_{\tau = 1}^{t} \frac {\Gamma_s^3(t)} {\Gamma_s^2(t)} = 0. \nonumber
\end{align}
\normalsize
\vspace{-0.4cm}

Unlike SBSs, the cloud has the knowledge on the demands over the whole network. Based on this global knowledge, cloud learns the caching strategy $\pi_c$ using the steps of \eqref{eq:Regret} by modifying the action vector to $\mathcal{A}_c = [\mathcal{A}_s]_{s \in \mathcal{S}}$, the corresponding utilities to $\Upsilon_c(\mathcal{A}_c) = \sum_{s = 1}^{\mathcal{S}} \Upsilon_s(\boldsymbol{\Xi}_s^{k_{s}}, \boldsymbol{\bar{g}}_s)$ and regret estimations to $\tilde{r}_{c}(\mathcal{A}_c) = \sum_{s = 1}^{\mathcal{S}} \tilde{r}_{s,\Xi_s^{k_s}}$.
\vspace{-0.3cm}
\subsection{Cache Eviction Algorithm}
\vspace{-0.1cm} 
To update the SBSs cache, existing contents need to be evicted due to the constrained cache size. For simplicity, we assume only a single content is evicted at time $t$. At every time $T_2$, each SBS observes the request for the cached contents. Based on the number of requests, each SBS builds the Gibbs-Sampling based distribution as:

\vspace{-0.4cm}
\begin{equation}
G_{sf}(t) = \frac {\exp(- \sum_{\tau=1}^{t-1} \pi_{sf}(\tau))} {\sum_{\forall f' \in \boldsymbol{\Xi}_s} \exp(- \sum_{\tau=1}^{t-1} \pi_{sf'}(\tau))}, \ \forall f \in \boldsymbol{\Xi}_s. 
\end{equation}
\vspace{-0.4cm}

From the above equation, the content with least popularity will be evicted from the cache. Using the Gibbs-Sampling based probability distribution, each SBS evicts the content and caches new content based on $\boldsymbol{\pi}'$ given by:
\vspace{-0.2cm}
\begin{equation}
\textstyle \boldsymbol{\pi}' = (1-\beta)\boldsymbol{\pi}_{s} + \beta \boldsymbol{\pi}_c,
\vspace{-0.2cm}
\end{equation}
\noindent where $\boldsymbol{\pi}_{s}$ and $\boldsymbol{\pi}_c$ represents the caching strategy at SBS $s$ and cloud respectively and $\beta$ captures the local/global tradeoff. Note that due to the assumption of time scale separation over three phases therein, the proposed solution does not assure global optimality of the network utility maximization.
	
\vspace{-0.2cm}
\section{Simulation Results}
\vspace{-0.2cm}
\begin{figure}[b]%
\centering
\vspace{-0.2cm}
\includegraphics[width = 7.25cm, height = 5.00cm]{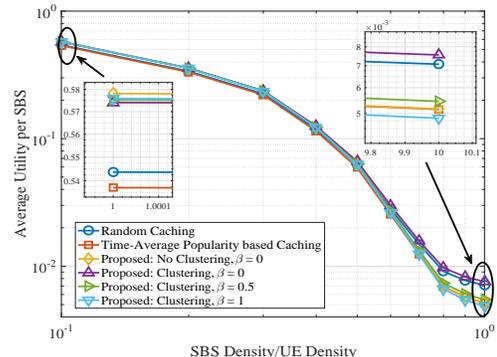}
\vspace{-0.3cm}
\caption{Average utility per SBS vs SBS Density/UE Density, $p_s $ $=$ $\mathrm{23dBm}$, $d$ $=$ $50$, $F$ $=$ $500$, $\xi_c = 0.0002$, $\xi_s = 0.01$, $\alpha = \beta$}
\vspace{-0.5cm}
\label{fig:SBSDensity}
\end{figure}%
In this section, we analyze the performance of the proposed mechanism and examine insights of the local/global tradeoff ($\beta$) under several deployment and caching scenarios. By assuming a system bandwidth of 1.4MHz, the performance of the proposed scheme is compared against two baseline schemes: random caching (B1) and time-average content popularity based caching (B2). Both baselines and proposed solution uses random RBs to serve users' requests. Further, $\frac {\lambda_{\mathrm{SBS}}} {\lambda_{\mathrm{UE}}} = 0.1$ denotes a sparse network while $\frac {\lambda_{\mathrm{SBS}}} {\lambda_{\mathrm{UE}}} = 1$ denotes a dense network. Fig. \ref{fig:SBSDensity} shows the per-SBS utility as a function of the ratio of SBS density to user density. With increased $\lambda_{\mathrm{SBS}}/\lambda_{\mathrm{UE}}$, the gains of the proposed scheme ($\beta = 0$) vary from 6\%-10\% and 8\%-40\% compared to B1 and B2, respectively. Meanwhile, the proposed scheme with clustering ($\alpha = \{0, 0.5\}$) achieves 23\%, 6\% 
gains over the proposed scheme without clustering. \\
\begin{figure}[t]%
\centering
\vspace{-0.3cm}
\includegraphics[width = 9.0cm, height = 5.45cm]{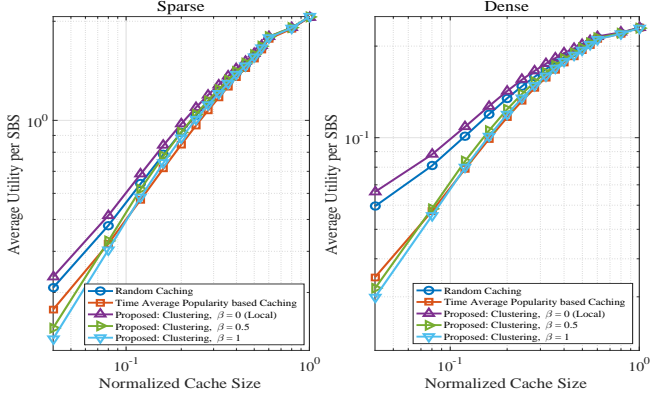}
\vspace{-0.8cm}
\caption{Average Utility vs cache size, $C_f = 50 \mathrm{Gbps}$, $\alpha = \beta$}
\vspace{-0.3cm}
\label{fig:CacheSize}
\end{figure}%
\indent Fig. \ref{fig:CacheSize} shows the variation of the per-SBS utility as a function of cache size. For a small cache size, the proposed scheme ($\beta = 0$) achieves \{10\%, 13\%\} and \{25\%, 56\%\} gains over baselines B1 and B2, respectively for \{sparse, dense\} scenarios. With the increasing cache size, the proposed scheme ($\beta = 0$) achieves 7\% and 28\% gains over baselines B1 and B2 for both scenarios. \\
\begin{figure}[b]%
\centering
\vspace{-0.3cm}
\includegraphics[width = 8.75cm, height = 4.75cm]{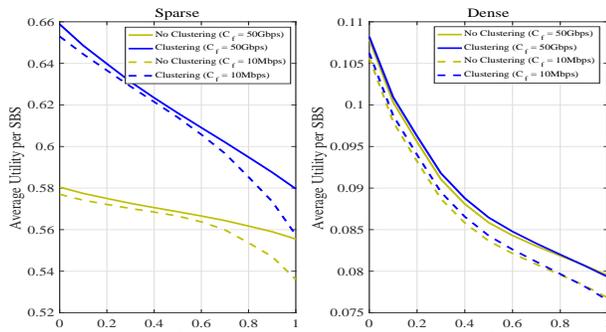}
\vspace{-1.0cm}
\caption{Local/global tradeoff for sparse/dense scenarios, $d$ $=$ $50$, $F$ $=$ $500$}
\vspace{-0.3cm}
\label{fig:CacheSize5}
\end{figure}%
\begin{figure}[t]%
\centering
\vspace{-0.3cm}
\includegraphics[width = 8.5cm, height = 5.50cm]{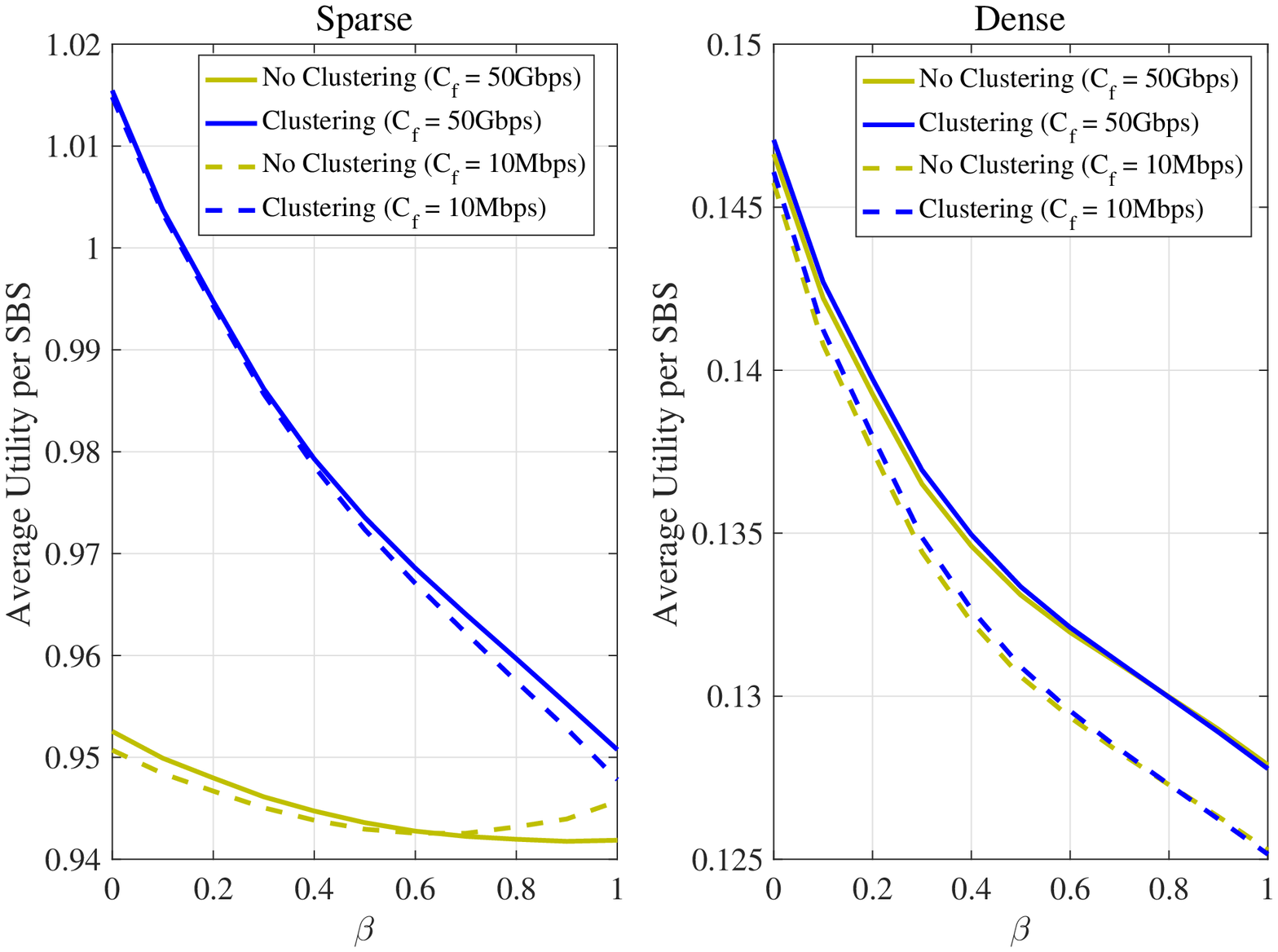}
\vspace{-0.4cm}
\caption{Local/global tradeoff for sparse/dense scenarios, $d$  $=$ $100$, $F$ $=$ $500$}
\label{fig:CacheSize10}
\vspace{-0.3cm}
\end{figure}%
\indent Fig. \ref{fig:CacheSize5} and \ref{fig:CacheSize10} shows the tradeoff between local and global learning. It can be observed that  the local clustering always performs better than no clustering approach for sparse and dense scenarios. At $\beta = 0.8$ for $\mathrm{dense}$ scenario, both schemes yield the same utility for small cache size. Further increasing $\beta$ makes the no clustering approach better than the local clustering. In addition, decreasing the fronthaul capacity has no impact of local/global tradeoff parameter. When the cache size increases, clustering approach is slightly better than non-clustering for dense scenario as shown in Fig. \ref{fig:CacheSize10}. Furthermore, at $\beta = 0.7$ for $\mathrm{dense}$ scenario, both schemes yield the utility. By increasing $\beta$ further makes the no clustering approach better than the local clustering. Furthermore, there is no impact of fronthaul capacity on $\beta$.
\vspace{-0.3cm}
\section{Conclusion} 
\vspace{-0.2cm}
In this letter, we investigated content caching in cloud-aided wireless networks, where SBSs store contents from a large content library. We proposed a clustering algorithm based on Gaussian similarity. Using the regret learning mechanism at the SBSs and the cloud, we proposed a per-SBS caching strategy that minimizes the service delay in serving users' requests. In addition, we investigated the tradeoff between local and global content popularity on the proposed algorithm for sparse and dense deployments.

\end{document}